# Dynamic balancing of super-critical rotating structures using slow-speed data via parametric excitation

Shachar Tresser, Amit Dolev and Izhak Bucher

Dynamics Laboratory, Technion, Israel

## Abstract

High-speed machinery is often designed to pass several "critical speeds", where vibration levels can be very high. To reduce vibrations, rotors usually undergo a mass balancing process, where the machine is rotated at its full speed range, during which the dynamic response near critical speeds can be measured. High sensitivity, which is required for a successful balancing process, is achieved near the critical speeds, where a single deflection mode shape becomes dominant, and is excited by the projection of the imbalance on it. The requirement to rotate the machine at high speeds is an obstacle in many cases, where it is impossible to perform measurements at high speeds, due to harsh conditions such as high temperatures and inaccessibility (e.g., jet engines).

This paper proposes a novel balancing method of flexible rotors, which does not require the machine to be rotated at high speeds. With this method, the rotor is spun at low speeds, while subjecting it to a set of externally controlled forces. The external forces comprise a set of tuned, response dependent, parametric excitations, and nonlinear stiffness terms. The parametric excitation can isolate any desired mode, while keeping the response directly linked to the imbalance. A software controlled nonlinear stiffness term limits the response, hence preventing the rotor to become unstable. These forces warrant sufficient sensitivity required to detect the projection of the imbalance on any desired mode without rotating the machine at high speeds. Analytical, numerical and experimental results are shown to validate and demonstrate the method.

# Nomenclature

| | |
|---|---|
| $a_j$ | Amplitude of the response of the $j^{th}$ mode |
| $A_j$ | Response of the $j^{th}$ mode |
| $\mathbf{C}$ | Damping and gyroscopic matrix |
| $\mathbf{D}$ | Damping matrix |
| $D_i$ | Differentiation operator w/r to time scales i |
| $\mathbf{f}_{ib}$ | Imbalance force vector |
| $\mathbf{f}_m$ | Modal imbalance force vector |
| $\mathbf{f}_{co}$ | Correction masses force vector |
| $\Delta f_m$ | Modal trial mass |
| $\Delta \mathbf{f}_{ib}$ | Trial mass vector |
| $\tilde{\mathbf{f}}_m$ | Modal imbalance force vector at trial run |
| $\mathbf{f}_{nl}$ | Nonlinear force vector |
| $\mathbf{G}$ | Gyroscopic matrix |
| $i$ | $\sqrt{-1}$ |
| $\mathbf{I}$ | Identity matrix |
| $k_p$ | Parametric excitation's stiffness (pumping amplitude) |
| $k_{pa,\min}$ | Minimal required pumping amplitude |
| $\mathbf{K_p}(t)$ | Time dependant stiffness matrix |
| $\mathbf{K}$ | Stiffness matrix |
| $k_3$ | Cubic stiffness constant |
| $\mathbf{M}$ | Mass matrix |
| $\mathbf{q}$ | Vector of degrees of freedom |
| $S_\bullet$ | Sensitivity of the response to $\bullet$ |
| $t$ | time |
| $\alpha$ | angular location of trial mass |
| $\beta_j$ | Response phase of the $j^{th}$ mode |
| $\Gamma$ | Modal stiffness matrix |
| $\varepsilon$ | Small non-dimensional number |
| $\phi_n$ | $n^{th}$ mode shape |
| $\mathbf{\Phi}$ | Mass normalized modal matrix |
| $\varphi$ | phase |
| $\mathbf{\eta}$ | vector of modal degrees of freedom |
| $\sigma$ | Detuning parameter |
| $\sigma_{opt}$ | Optimal detuning parameter |
| $\omega_n$ | Natural frequency of the $n^{th}$ mode |
| $\Omega$ | Speed of rotation |
| $\psi_j$ | Response phase of the $j^{th}$ mode |
| $\zeta_n$ | Damping ratio of the $n^{th}$ mode |

# 1  Introduction

The main cause for vibration in rotating structures is "imbalance" which is a common term to describe the effect of minute manufacturing imperfections and deviations of the mass center from the rotation axis. While the structure is rotating, the imbalance gives rise to rotating forces whose effect on individual modes of vibration is proportional to the projection of the imbalance axial distribution on each mode. The structure's response is composed of a superposition of all mode shapes (eigenvectors), where indeed each mode is excited by the projection of the imbalance on the individual mode [1,2].

Imbalance is routinely compensated for by adding (or removing) small correction masses to the structure at pre-defined axial locations. These corrective masses are placed such that their radial and angular locations eliminate the effect of imbalance on all the vibration modes within the relevant speed range. These corrective masses are computed solely from measured vibrations during operation in a so called "balancing process" [3–6]. High speed rotors are usually balanced using either the "Influence Coefficient Method", "Modal Balancing" or the "Unified Balancing Approach" [3–11]. The calculation of the correction masses using the aforementioned procedures is based on measuring the imbalance response close to critical speeds, where the vibration levels and sensitivity are sufficiently high.

Usually, the balancing procedure requires the rotor to be spun at the entire anticipated operating speed range during normal service [5].

The requirement to spin the structure at its entire operating speed range is a major obstacle in many cases. Frequently, reaching high rotation speeds involves conditions that do not enable measurements of the imbalanced response (e.g., jet engines where operating conditions involve very high temperatures and hazardous environmental conditions surrounding the rotor). The technical challenges often lead to one of the following:

a. A conservative over-design, trying to keep the critical speeds well above the operating speed.
b. A compromise on the balancing procedure, by using commercial balancing machines [6][12][13], which are incapable of rotating at sufficiently high speeds (e.g., small Jet engines rotate at 100,000 rev/min, while balancing machines are normally limited to 3000 rev/min.). A commercial balancing machine calculates two correction masses that cancel the reaction forces while spinning the rotor at a low speed, assuming that the rotor is rigid [3–6]. Although rigid rotor balancing is a very simple and straightforward procedure, it cannot identify the projection of the imbalance on high speed flexible modes. In fact, in some cases rigid rotor balancing can even increase the projection of the imbalance on high speed related flexible modes [5,11].
c. Damping elements (e.g., squeeze film, magnetic [14,15]) are proposed as a common design alternative for poorly balanced rotors, these add weight and often unacceptable complexity.

Given these limitations, a different approach is required, which substantially amplifies the projection of the imbalance on high frequency modes (modes related to high rotation speeds), while rotating at low speeds. To achieve this goal, dual frequency parametric excitation techniques are adopted here.

Parametric amplifiers are known for their high amplification and selectivity and indeed these find use in various fields of physics and engineering. In recent years, parametric resonators (PR) have found use in various engineering areas [16–19], mainly due to Micro- and Nano Electro Mechanical Systems technologies, which allow low fabrication costs, good performances [20], and easy integration into engineering systems [18,21]. Usually, the principal parametric resonance is employed where the first linear instability tongue in the

Ince-Strutt diagram [22–26] resides. Parametric excitation can amplify some forces through a combination of their frequencies in a manner the structural dynamics favours [27–29].

Recently, Dolev and Bucher introduced a tuneable parametric amplifier (PA) with a hardening, Duffing-type nonlinearity [30,31]. The PA differs from what has been presented so far because it frees the amplified signal frequencies from being an integer multiple of the natural frequency [32], as commonly assumed under parametric resonance. The proposed approach does not pose an impractical constraint as done in some cases, yet it obtains a better performance than an alternative mixing approach [33] which relaxes this condition, but does not produce sufficient amplification without very high pumping levels [33,34] (high excitation forces). It was shown that the PA significantly amplified a selected mode of vibration, even when the external force frequency was much lower than the system's natural frequency.

The present paper embraces the basics of the method in [31] by employing a dual frequency parametric excitation as a mean to detect the projection of the imbalance on any desired mode, while rotating much slower than the critical speeds.

The paper outlines the proposed method with a brief introduction of the relevant dynamics and the governing equations. Later, both numerical simulations and experimental results are shown to validate the analytical model, and the novel balancing method.

## 2  Proposed Method

In order to describe the proposed approach, it is necessary to define basic terms from the dynamics of rotating structures. Indeed, the response to imbalance of a linear structure is the basic layer on which the method is built. The equations of motion for a constant speed of rotation, can be written in complex vector format as:

$$\mathbf{M}\ddot{\mathbf{q}} + \mathbf{C}\dot{\mathbf{q}} + \mathbf{K}\mathbf{q} = \Omega^2 \mathbf{f}_{ib} e^{i\Omega t}, \tag{1}$$

where $\mathbf{C} = \mathbf{D} + \Omega \mathbf{G}$. Here $\mathbf{M}$, $\mathbf{D}$, $\mathbf{G}$ and $\mathbf{K}$ are the mass, damping, gyroscopic and stiffness matrices respectively, $\mathbf{f}_{ib}$ is the imbalance force vector, representing its spatial distribution along the structure, $\mathbf{q}$ is the degrees of freedom vector, and $\Omega$ the rotation speed. For simplicity, in this work, both the experimental system and the analysis ignore the gyroscopic effect and the change of mode shapes with speed of rotation.

The force exerted by imbalance depends on the speed of rotation, $\Omega$. The response to imbalance, when the gyroscopic effect is negligible, is composed of a superposition of all mode shapes, where each mode is excited by the projection of the imbalance on it, as shown below [1]:

$$\mathbf{q}(z,t) = \sum_{n=1}^{\infty} \frac{\boldsymbol{\phi}_n \left(\boldsymbol{\phi}_n^T \mathbf{f}_{ib}\right) \Omega^2 e^{i\Omega t}}{\omega_n^2 - \Omega^2 + i2\zeta_n \omega_n \Omega}, \qquad (2)$$

here $\boldsymbol{\phi}_n$, $\zeta_n$ and $\omega_n$ are the $n^{th}$ mode shape, damping ratio and natural frequency, respectively. Clearly, $\boldsymbol{\phi}_n^T \mathbf{f}_{ib}$ controls the amplitude of vibration of each mode of vibration. One can learn from Eq.(2) that any set of correcting masses, yielding a generalized complex force amplitude $\mathbf{f}_{co}$ such that

$$\boldsymbol{\phi}_n^T \mathbf{f}_{ib} + \boldsymbol{\phi}_n^T \mathbf{f}_{co} \approx 0 \qquad (3)$$

eliminates the source of vibration affecting mode $n$. Indeed, a common practice, as mentioned in the introduction [3-11], is to spin the system close to the critical speed of each mode at a turn, increasing the speed of rotation to $\Omega \approx \omega_n$, then one can assess the required correction masses that would comply with Eq.(3).

## 2.1 Derivation of the proposed method through an example

Consider the system described in Fig.1, where a rigid rotor is mounted on a plate, free to move only in the horizontal plane, and subjected to two externally controlled forces for which:

$$f_i = -\left\{k_{pa,i} \cos(\omega_a t - \phi_a) + k_{pb,i} \cos(\omega_b t - \phi_b)\right\} x_i - k_{3,i} x_i^3, \quad i = 1, 2. \qquad (4)$$

Here, $k_{pa,i}$, $\omega_a$, $\phi_a$, $k_{pb,i}$, $\phi_b$ and $k_{3,i}$ are tunable parameters and $x_i$ is the measured displacement of the structure where $f_i$ is applied.

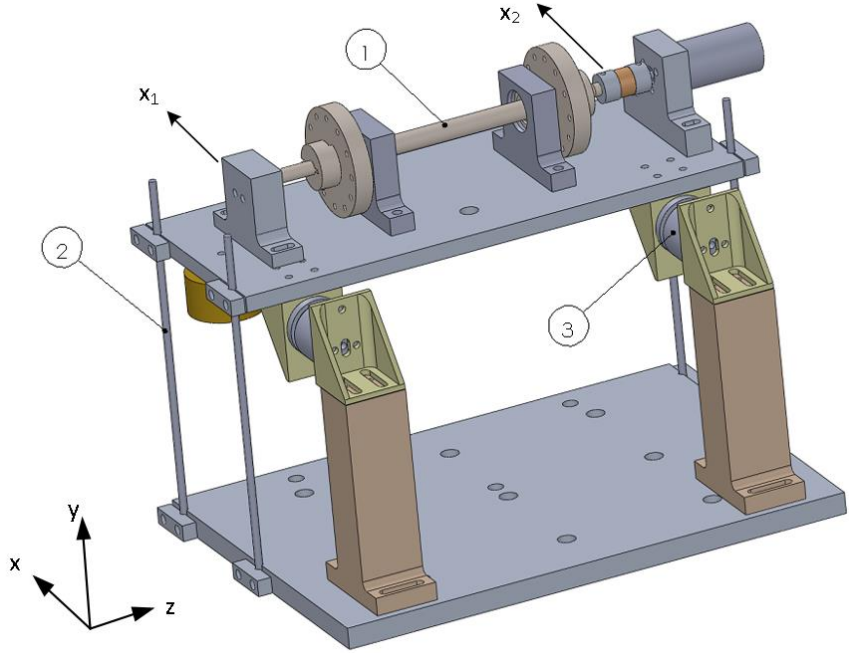

*Figure 1 – Rigid rotor balancing demonstrator, consisting of rotating shaft and discs (1), flexible foundation supports (2) and electromechanical actuators (3) inducing the vibration-dependent parametric excitation*

The equations of motion of the system in figure 1, including the effect of Eq.(4) are:

$$\mathbf{M}\ddot{\mathbf{q}} + \mathbf{C}\dot{\mathbf{q}} + \left(\mathbf{K} + \mathbf{K_p}(t)\right)\mathbf{q} + \mathbf{f}_{nl}(\mathbf{q}) = \Omega^2 \mathbf{f}_{ib}. \tag{5}$$

Here $\mathbf{K_p}(t)$ and $\mathbf{f}_{nl}$ are a time dependant stiffness matrix and the nonlinear force vector, produced by the controlled forces shown in Eq.(4). By assuming low damping, small nonlinearity and small parametric excitation amplitudes, and introducing a small non-dimensional number $0 < \varepsilon \ll 1$, the equations of motion can be transformed into a form with several diagonal matrices:

$$\ddot{\boldsymbol{\eta}} + \boldsymbol{\Gamma}\boldsymbol{\eta} + \varepsilon\left(\mathbf{C_m}\dot{\boldsymbol{\eta}} + \mathbf{K_{p_m}}(t)\boldsymbol{\eta} + \mathbf{f_{nl}}_m\right) = \Omega^2 \mathbf{f_m}, \tag{6}$$

where the subscript *m* denotes matrices and vectors expressed in modal coordinates. The transformation to modal coordinate is given by:

$$\mathbf{q} = \boldsymbol{\Phi}\boldsymbol{\eta}, \tag{7}$$

where $\boldsymbol{\Phi}$ is the non-rotating mass normalized modal matrix, obtained by solving

$$\mathbf{M\Phi\Gamma} = \mathbf{K\Phi}, \quad \mathbf{\Phi} \in \mathbb{R}^{N \times N}, \quad \mathbf{\Gamma} = \operatorname*{diag}_{i=1..N}\{\omega_i^2\}. \tag{8}$$

The equations of motion, Eq.(6), are solved by using the following, standard, multiple scales expansion [30,31,35]:

$$\mathbf{\eta}(t) \approx \mathbf{\eta_0}(t_0, t_1) + \varepsilon \mathbf{\eta_1}(t_0, t_1), \quad t_i = \varepsilon^i t. \tag{9}$$

We denote $D_i$ as differentiation with respect to time scale $t_i$.

$O(\varepsilon^0)$:
$$\left(\mathbf{I}D_0^2 + \mathbf{\Gamma}\right)\mathbf{\eta_0} = \Omega^2 \mathbf{f}_m. \tag{10}$$

$O(\varepsilon^1)$:
$$\left(\mathbf{I}D_0^2 + \mathbf{\Gamma}\right)\mathbf{\eta_1} = -\left(\mathbf{C_m} D_0 \mathbf{\eta_0} + \mathbf{K_{p_m}} \mathbf{\eta_0} + \mathbf{f_{nl_m}}\right) - 2\mathbf{I} D_0 D_1 \mathbf{\eta_0}. \tag{11}$$

The solution of Eq.(10) consists of vibration at all of the natural frequencies (the homogenous solution), and at the spin frequency (the private solution):

$$\eta_{j0} = A_j(t_1)e^{i\omega_j t_0} + \frac{\Lambda_j}{2}e^{i\Omega t_0} + CC, \quad \Lambda_j = \frac{\Omega^2 f_{mj} e^{-i\varphi_j}}{(\omega_j^2 - \Omega^2)}, \tag{12}$$

where $\omega_j$ is the $j^{th}$ natural frequency and CC stands for complex conjugate of the preceding terms. Substituting Eq.(12) into Eq.(11) leads to:

$$\left(\mathbf{I}D_0^2 + \mathbf{\Gamma}\right)\mathbf{\eta_1} = \text{RHS}. \tag{13}$$

Where RHS is the right hand side of Eq.(11), consisting of the following terms:

a. Terms arising from the damping and the differentiation with respect to time (first and last terms in RHS of Eq.(11). These terms lead to the decay of the homogenous solution due to damping when no other secular terms are present (terms at the natural frequency).

b. Terms arising due to nonlinearity. These terms are the result of $\eta_{jo}^3$ (see Eq. (12)), and they consist of the following harmonic terms:

$$\left(\alpha_1 e^{i\Omega t_0} + \alpha_2 e^{-i\Omega t_0} + \alpha_3 e^{i\omega_1 t_0} + \alpha_4 e^{-i\omega_1 t_0} + \alpha_5 e^{i\omega_2 t_0} + \alpha_6 e^{-i\omega_2 t_0}\right)^3, \tag{14}$$

where $\alpha_{1-6}$ are constant coefficients.

c. Terms arising due to the parametric excitation, which have the following form:

$$\sum_{\substack{p=a,b \\ j=1,2}} \Lambda_j c_{pj} \exp\left(i\left(\pm\Omega\pm\omega_p\right)t \mp \varphi_p \mp \varphi_j\right) + \sum_{\substack{p=a,b \\ j=1,2}} c_{pj} \exp\left(i\left(\pm\omega_j \pm \omega_p\right)t \mp \varphi_p\right), \tag{15}$$

where $c_{pj}$ are coefficients, the subscript – $p$ stands for the pumping frequency (there are two here, $\omega_a$ and $\omega_b$), and $j$ is the indication of quantities linked to the $j^{th}$ natural frequency related to the appropriate mode.

In order to excite the $n^{th}$ natural frequency, the parametric frequencies are set to:

$$\omega_a = 2\omega_n + \varepsilon\sigma_a, \quad \omega_b = \omega_n - \Omega + \varepsilon\sigma_b, \tag{16}$$

where $\sigma_i$ is a detuning parameter. The first frequency ($\omega_a$) produces principal parametric resonance, whose role is to significantly amplify the response near the desired natural frequency. The latter ($\omega_b$) is referred to as the "blending element", whose role is to couple the response to the imbalance, such that their sum of frequencies corresponds to the natural frequency of the mode to be balanced.

Since we wish to maximize the effect of the blending element, we require that no other secular terms will arise due to the non-linearity (so that there will be no subharmonic or superharmonic resonances). By expanding Eq. (14) the requirement above means that none of these combination should exist:

$$\omega_j \neq \{3\omega_k, \ 2\omega_k \pm \omega_j, \ 2\omega_k \pm \Omega, \ 2\Omega \pm \omega_k, \ 3\Omega\}, \quad j,k = 1,2. \tag{17}$$

The nonlinearity is required in order to limit the response, and achieve a steady state solution [35]. As shown in Eq.(12), the response to imbalance is a superposition of the projection of the imbalance on all modes, hence secular terms arise due to the projection of the imbalance on all modes, as can be seen in Eq. (15). Since only the projection on the $n^{th}$ mode is of interest, it proves to be useful to set the parametric excitation so it produces secular terms only at the $n^{th}$ natural frequency. This is achieved by setting the modal time-dependent stiffness matrix $\mathbf{K}_{P_m}(t)$ to be diagonal. The case where $\mathbf{K}_{P_m}(t)$ is not diagonal is addressed in Appendix A.

The matrix $\mathbf{K}_{p_m}(t)$ is of the form:

$$\mathbf{K}_{p_m}(t) = \mathbf{\Phi}^T \mathbf{K}_p(t) \mathbf{\Phi} = \begin{bmatrix} k_{p1}(t)\phi_{11}^2 + k_{p2}(t)\phi_{21}^2 & k_{p1}(t)\phi_{11}\phi_{12} + k_{p2}(t)\phi_{21}\phi_{22} \\ k_{p1}(t)\phi_{11}\phi_{12} + k_{p2}(t)\phi_{21}\phi_{22} & k_{p1}(t)\phi_{12}^2 + k_{p2}(t)\phi_{22}^2 \end{bmatrix}. \tag{18}$$

Hence, the requirement is that the parametric excitations must obey:

$$k_{p1}(t)\phi_{11}\phi_{12} + k_{p2}(t)\phi_{21}\phi_{22} = 0. \tag{19}$$

To achieve this, we set

$$k_{pb1} = k_{pb}, \quad k_{pb2} = \frac{-\phi_{11}\phi_{12}}{\phi_{21}\phi_{22}} k_{pb}, \quad \varphi_{a1} = \varphi_{a2} = \varphi_a, \quad \varphi_{b1} = \varphi_{b2} = \varphi_b. \tag{20}$$

We transform to polar representation, as commonly done in similar cases [35], by setting:

$$A_j(t_1) = \frac{1}{2} a_j(t_1) e^{i\beta_j(t_1)}. \tag{21}$$

Substituting Eq.(16),(20) and (21) to Eq.(13), and equating secular terms to zero to ensure periodic response and avoid divergence, leads to a system of two complex equations. Separating to real and imaginary parts generates a system of four equations with four unknowns: the amplitudes and phases $a_i(t_1)$, $\beta_i(t_1)$, $i = 1, 2$.

For the case where the $j^{th}$ mode is to be excited, the equations of the $k^{th}$ mode (the unexcited mode) are:

$$\begin{aligned}\Im: \quad & 2\omega_k a_k' + 2\omega_k^2 \zeta_k a_k = 0 \\ \Re: \quad & 1/2 a_k \begin{pmatrix} 6\Lambda_1 \Lambda_2 \cos(\varphi_1 - \varphi_2)\left(k_{32}\phi_{2k}^3 \phi_{2j} + k_{31}\phi_{1k}^3 \phi_{1j}\right) + \\ +12 a_j^2 \left(k_{32}\phi_{22}^2\phi_{21}^2 + k_{31}\phi_{12}^2\phi_{11}^2\right) + 3\Lambda_k^2 \left(k_{31}\phi_{1k}^4 + k_{32}\phi_{2k}^4\right) + \\ +3\Lambda_j^2 \left(k_{31}\phi_{12}^2\phi_{11}^2 + k_{32}\phi_{22}^2\phi_{21}^2\right) - 4\omega_k \beta_k' + 6 a_k^2 \left(\phi_{2k}^4 k_{32} + \phi_{1k}^4 k_{31}\right) \end{pmatrix} = 0 \end{aligned} \tag{22}$$

Where $\bullet'$ represents differentiation with respect to time scale $t_1$. As can be seen from the imaginary part of Eq.(22), the unexcited mode decays with time due to damping. The equations of the $j^{th}$ mode are:

$$\Im: \begin{aligned}&\Lambda_j \sin(\sigma_b t_1 - \beta_j - \varphi_j - \varphi_b) k_{pb} \phi_{1j} |\Phi| + \sin(\sigma_a t_1 - 2\beta_j - \varphi_a) a_j k_{pa} \phi_{1j} |\Phi| + \\ &\pm 4\omega_j \phi_{2k} (\omega_j \zeta_j a_j + a'_j) = 0\end{aligned} \quad (23)$$

$$\Re: \begin{aligned}&\pm 12 a_j \Lambda_1 \Lambda_2 \cos(\varphi_1 - \varphi_2) \phi_{2k} (k_{32} \phi_{22} \phi_{21}^3 + k_{31} \phi_{11}^3 \phi_{12}) \mp 8 a_j \phi_{2k} \omega_j \beta'_j + \\ &\pm 3 a_j^3 \phi_{2k} (\phi_{11}^4 k_{31} + \phi_{21}^4 k_{32}) + 2\Lambda_j \cos(-\sigma_b t_1 + \beta_1 + \varphi_1 + \varphi_b) k_{pb} \phi_{1j} |\Phi| + \\ &+ 2 a_j \cos(\sigma_a t_1 - 2\beta_j - \varphi_a) k_{pa} \phi_{1j} |\Phi| \pm 24 a_j a_k^2 \phi_{2k} (k_{31} \phi_{11}^2 \phi_{12}^2 + k_{32} \phi_{22}^2 \phi_{21}^2) + \\ &\pm 6 a_j \phi_{2k} \left[ \Lambda_j^2 (\phi_{1j}^4 k_{31} + \phi_{2k}^4 k_{32}) + \Lambda_k^2 (k_{31} \phi_{11}^2 \phi_{12}^2 + k_{32} \phi_{22}^2 \phi_{21}^2) \right] = 0\end{aligned} \quad (24)$$

where $|\Phi| = \phi_{11}\phi_{22} - \phi_{12}\phi_{21}$, $\pm$ is (+) for $j=1$, and (−) for $j=2$, $\mp$ is (−) for $j=1$, and (+) for $j=2$. Note that the expressions are similar for both modes since $\phi_{21} \approx -\phi_{22}$, as shown in section 3.1.

It is possible to convert these equations to an autonomous system by substituting:

$$\psi_{ja}(t_1) = -\sigma_a t_1 + 2\beta_j(t_1), \quad \psi_{jb}(t_1) = -\sigma_b t_1 + \beta_j(t_1). \quad (25)$$

At steady state, when $a'_j = \psi'_j = 0$, the amplitudes and phases are denoted by $a_{j0}, \psi_{j0}$, and it is required that:

$$\sigma_a = 2\sigma_b = 2\sigma, \quad \psi_{ja} = 2\psi_{jb} = 2\psi_j. \quad (26)$$

Substituting $a_k = 0$, Eq.(25) and Eq.(26) into Eq.(24) leads to the following equations:

$$a'_j = \frac{\sin(2\psi_j + \varphi_a) a_j k_{pa} \phi_{1j} |\Phi| + \Lambda_j \sin(\psi_j + \varphi_j + \varphi_b) k_{pb} \phi_{1j} |\Phi| \mp 4\omega_j^2 \zeta_j a_j \phi_{2k}}{\pm 4\omega_j \phi_{2k}}$$

$$\psi'_j = \frac{1}{\pm 8 a_j \omega_j \phi_{2k}} \begin{pmatrix} 2a_j \cos(2\psi_j + \varphi_a) k_{pa} \phi_{1j} |\Phi| + 2\Lambda_j \cos(\psi_j + \varphi_j + \varphi_b) k_{pb} \phi_{1j} |\Phi| + \\ \pm 12 a_j \Lambda_1 \Lambda_2 \phi_{2k} \cos(\varphi_1 - \varphi_2)(k_{31} \phi_{1j}^3 \phi_{1k} + k_{32} \phi_{2j}^3 \phi_{2k}) + \\ \pm 3 a_j^3 \phi_{2k} (k_{31} \phi_{1j}^4 + k_{32} \phi_{2k}^4) \mp 8 a_j \phi_{2k} \omega_j \sigma + \\ \mp 6 a_j \phi_{2k} (\Lambda_j^2 (k_{31} \phi_{1j}^4 + k_{32} \phi_{2j}^4) + \Lambda_k^2 (k_{31} \phi_{11}^2 \phi_{12}^2 + k_{32} \phi_{22}^2 \phi_{21}^2)) \end{pmatrix} \quad (27)$$

According to Eq.(27), the steady state amplitude of the excited mode (denoted by the subscript $j$) is given by:

$$a_{j0}(t_1) = \frac{-\Lambda_j k_{pb} \phi_{1j} |\Phi| \sin(\psi_{j0}(t_1) + \varphi_j + \varphi_b)}{k_{pa} \phi_{1j} |\Phi| \sin(2\psi_{j0}(t_1) + \varphi_a) \mp 4\omega_j^2 \zeta_j \phi_{2k}} \tag{28}$$

Isolation of $\psi_{j0}$ from the 2nd equation of Eq. (27) in closed form is not possible, and results in a transcendental equation, hence the phase can be computed by iterative search. An alternative approach has been proposed in [30] where the phase can be computed from the roots of a polynomial equation without iterations.

Frequency response curves and a comparison of the analytical model to numerical simulations are shown in Fig.2, for both modes. The analytical solutions are in excellent agreement with the numerically computed ones. Figure 2 shows that upon tuning the frequency of the parametric excitation, one can obtain large amplification of the vibratory response at either mode1 or mode 2. It is shown in Fig.2 that the response has multiple solutions, with 2 stable possible solutions (calculation of stability is done by applying a standard local stability check [35] as shown in appendix B). As will be explained later on in this section, the zone where a single solution exists is of greater interest. The response in case of a detuning leading to two stable solutions is addressed in section 3.2.1.

We point out three important conclusions from Eq.(28):

a. The response level depends only on the projection of the imbalance on the desired mode.

b. The magnitude of the response strongly depends on the phases of the modal imbalance $\varphi_j$ and the blending excitation's phase $\varphi_b$. Moreover, there is a simple relationship between these phases and a combination which gives rise to zero amplitude is:

$$\psi_{j0} + \varphi_j + \varphi_b = \{0, \pi\}. \tag{29}$$

c. The minimal pumping amplitude required, noted by $k_{pa,\min}$, is the value to nullify the denominator of Eq.(28), which denotes the linear stability threshold, as given by Eq. (30):

$$k_{pa,\min} \geq \frac{4\zeta_j \phi_{2k} \omega_j^2}{\phi_{1j} |\Phi|} \tag{30}$$

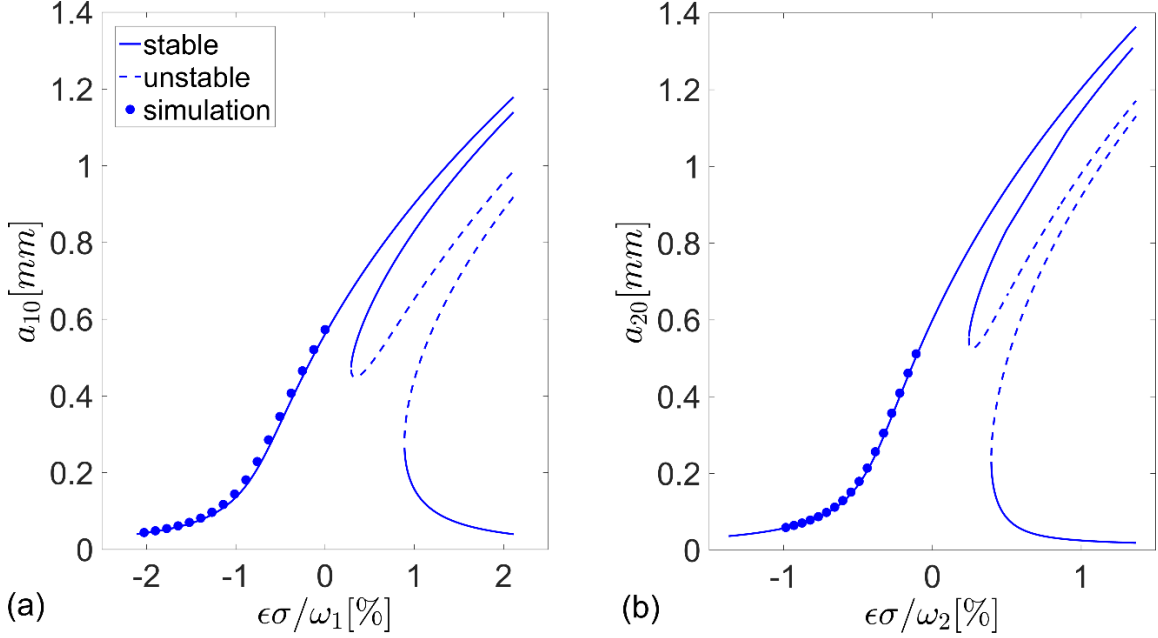

*Figure 2: Frequency response curves. (a) 1st mode ($a_{10}$); (b) 2nd mode ($a_{20}$). Numerical simulation (dots), analytical stable solutions (continuous lines), and unstable solutions (dashed lines).*

Since the response $a_{j0}(t_1)$ is periodic with respect to $\varphi_b$ and $\varphi_j$ (see Eq.(28)), sweeping through the former is the key to the proposed balancing procedure, as will be explained thoroughly in section 2.2. An effective balancing procedure requires two conditions:

a. Sufficiently large signals (high amplification) such that the measured amplitudes are well above the noise level.
b. Sensitivity to the imbalance and to the phase of the parametric excitation is sufficient to reduce the effect of noise and other dynamical effects to an acceptable level.

The sensitivities are defined as:

$$S_\Lambda = \frac{\partial a_j}{\partial \Lambda},\ S_\varphi = \frac{\partial a_j}{\partial \varphi}. \tag{31}$$

The sensitivities to the magnitude and phase of the imbalance are shown in Figure 3 (only for the 1st mode). Note that high sensitivities are obtained for detuning values that lead to a single valued solution (see also Figure 2).

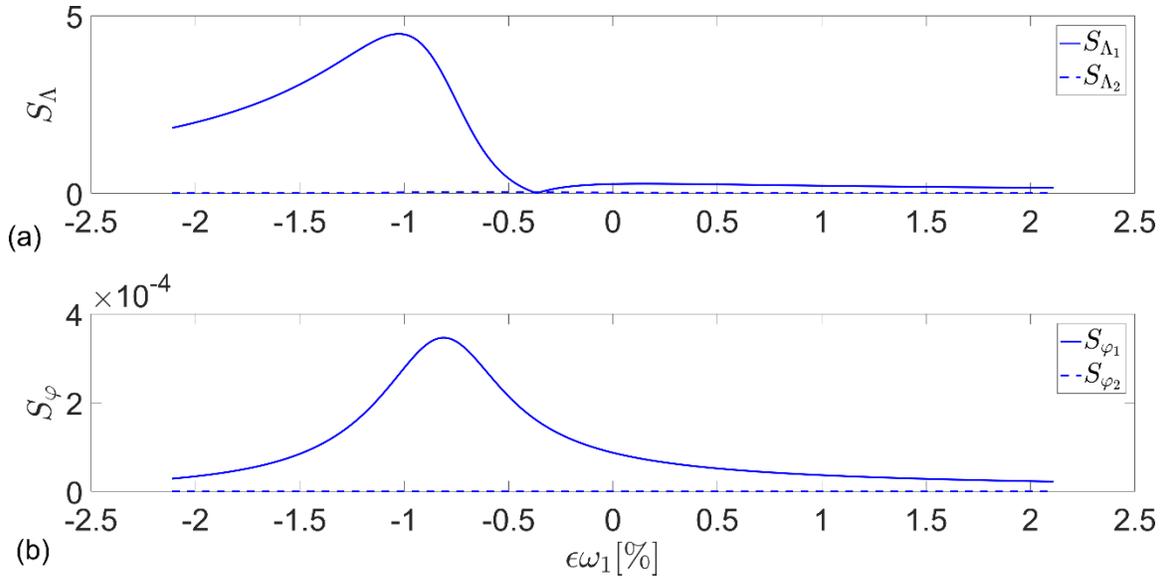

*Figure 3: Sensitivities of the system vs. the detuning, as computed for mode #1. (a) Sensitivity to the magnitude of the modal imbalance; (b) Sensitivity to the phase of the modal imbalance*

As can be seen from Figure 3, although the nonlinearity couples the response of all the modes, the steady state response is not sensitive to the projection of the imbalanced on the other (unexcited) mode.

## 2.2 Balancing procedure

The practical meaning of the three stated conclusions derived from Eq.(28), is that by setting the parametric excitation to be sufficiently large, according to Eq.(30), one can modify the value of the parametric excitation's phase, $\varphi_b$, until minimal amplitude is reached, denoted $\varphi_{b0}$ from now on. This allows one to find the location of the modal imbalance $\varphi_j$ according to Eq.(29). The phase of the modal imbalance $\varphi_j$ has two possible solutions spaced 180 degrees apart. The true location and magnitude of the imbalance can be found by placing a trial mass, resulting in a modal trial mass of $\Delta f_m$ (the projection of the trial mass on the mode). An alternative is to place a modal trial mass set [3], denoted by $\Delta \mathbf{f}_{ib}$, namely placing several trial masses corresponding to an individual mode. Instead of placing trial masses, one can apply equivalent synchronous forces by the voice coils, without the need for applying real trial masses, which may in some cases require disassembly and re-assembly of machine parts. The total modal imbalance at the trial run is:

$$\left|\tilde{f}_m\right|e^{-i\tilde{\varphi}_j} = \left|\tilde{f}_m\right|e^{-i\varphi_j} + \Delta f_m e^{-i(\varphi_j+\alpha)} \,. \tag{32}$$

The sweeping process of $\varphi_b$ is performed again, so that the phase $\tilde{\varphi}_j$ is found (again, two possible solutions 180 degrees apart). The modal trial mass should be placed about 90 degrees apart from $\varphi_j$ to achieve maximum change in $\varphi_j$. Since the imbalance at the trial run must lie between $\varphi_j$ and $\alpha$, the true location of $\varphi_j$ (and $\tilde{\varphi}_j$) can be found. The magnitude of the modal imbalance can be computed by the following expression:

$$f_m = \frac{|\tilde{f}_m|\sin(\pi - \alpha - \Delta\varphi_j)}{\sin(\Delta\varphi_j)}, \tag{33}$$

as shown in Figure 4 and Figure 5.

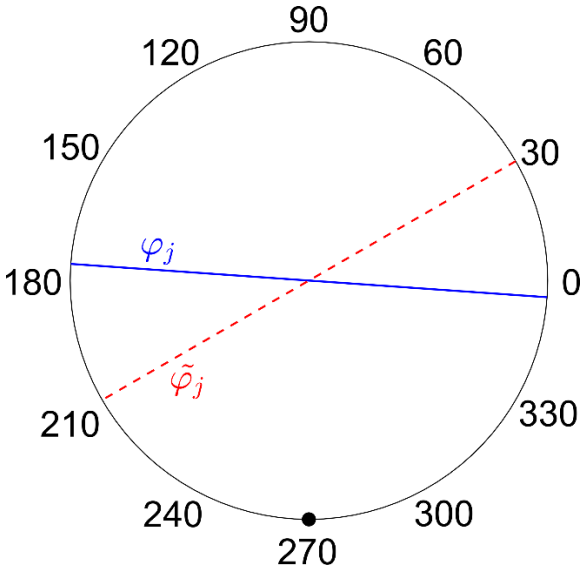

*Figure 4: possible solutions for $\varphi_j$ in the first run (blue line), and the trial run (red dashed line) with a trial mass (black dot).*

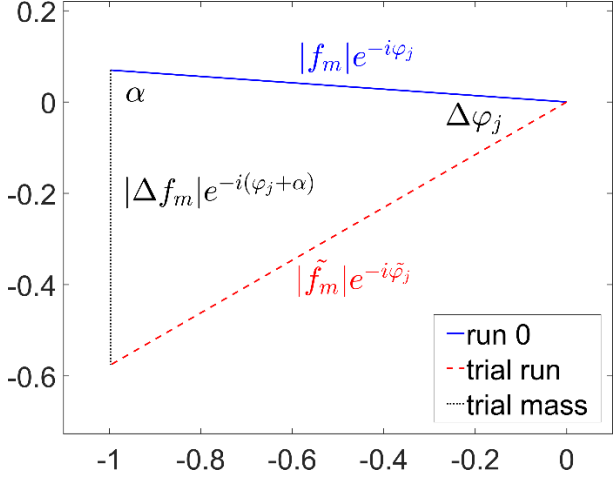

Figure 5: extraction of imbalance. First run (full line), trial run (dashed line) and trial mass (dotted line).

## 2.3 Sensitivity to various parameters

The balancing procedure is based on two principles: amplification of the response, and sensitivity to the phase of the excitation, denoted by $S_{\varphi b}$. A successful balancing procedure, requires proper selection of the various parameters, such as the excitations' levels $k_{pa}, k_{pb}$, the nonlinearity $k_3$, and the amount of detuning $\sigma$. As seen from Eq.(28), the amplification grows with an increase in the excitations' levels $k_{pa}, k_{pb}$. The magnitude of the blending element is $k_{pb}$, hence the sensitivity to the imbalance increases with it , as shown in

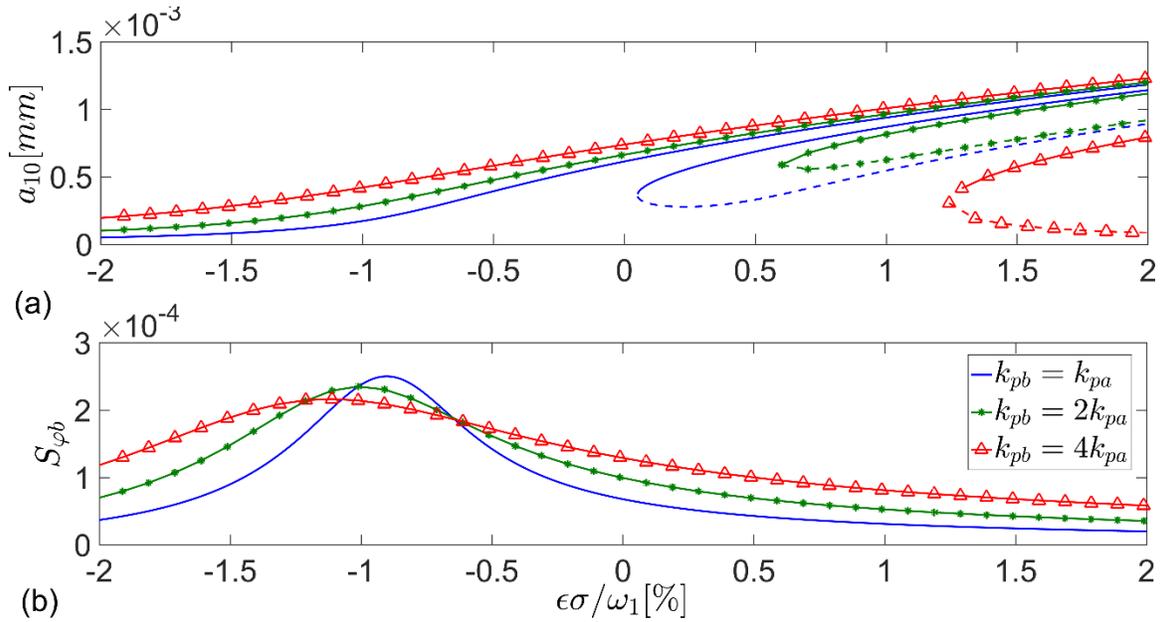

Figure 6. However $k_{pa}$, governs the principal parametric resonance, so it can greatly increase the amplitude, but causes the response to be less sensitive to the imbalance, even if $k_{pb}$ is increased as well (i.e., the sensitivity becomes narrow banded, as shown in Figure 7).

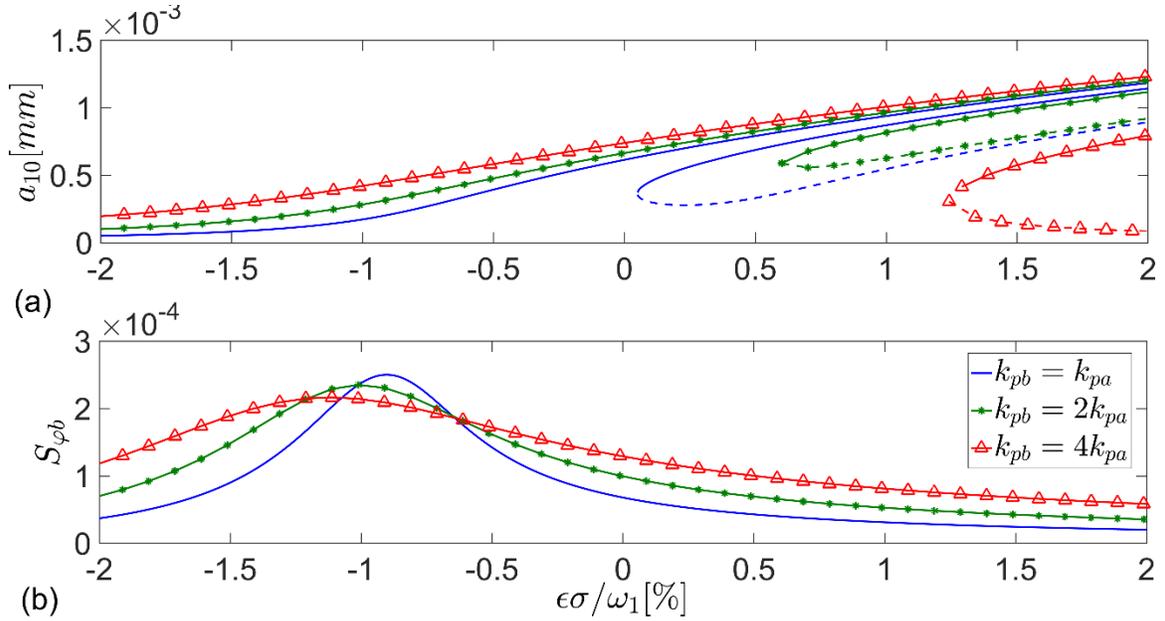

Figure 6: The effect of $k_{pb}$ on the amplitude (a) and sensitivity (b), stable solutions (continuous lines), and unstable solutions (dashed lines).

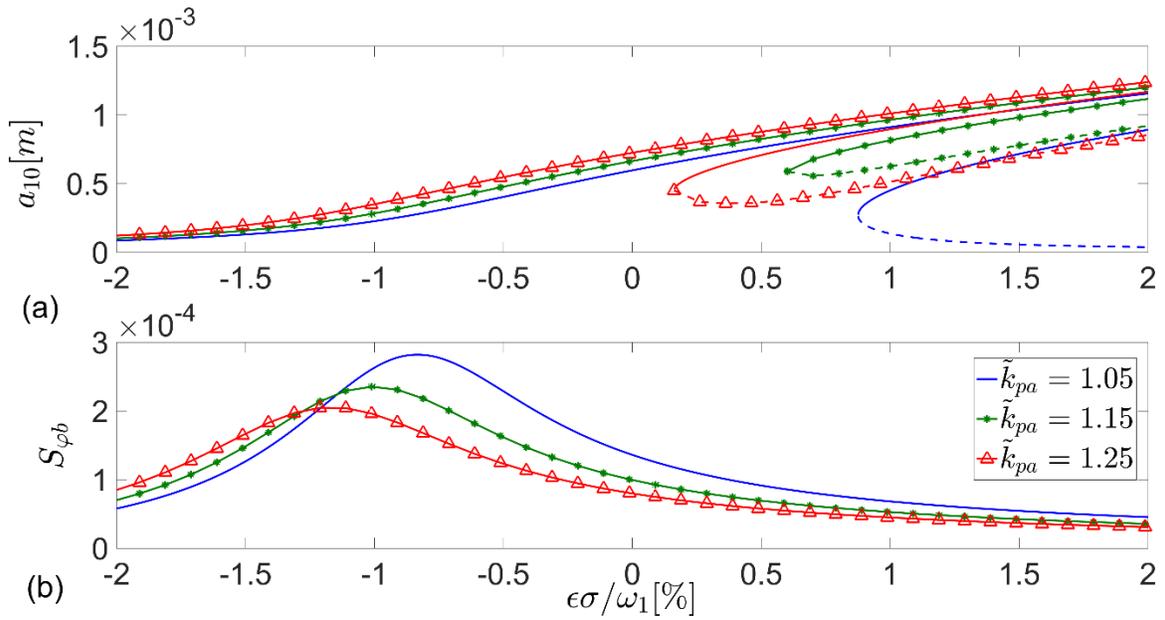

Figure 7: The effect of $\tilde{k}_{pa}$ ( $= k_{pa} / k_{pa,\min}$ ) on the amplitude (a) and sensitivity (b), for $k_{pb} = 2k_{pa}$, stable solutions (continuous lines), and unstable solutions (dashed lines).

The ideal detuning level, denoted by $\sigma_{opt}$, is the one for which the sensitivity to phase of the excitation is highest. This value can be found experimentally, since it is located at the point in which the frequency response curve starts bending upwards sharply, as can be seen in Figure 7 to Figure 10.

As mentioned earlier, the nonlinearity $k_3$ determines the maximal response amplitude at steady state. The nonlinearity should be set to achieve the desired steady state amplitude. Note that at $\sigma_{opt}$, where the sensitivity is the greatest, the nonlinearity has little effect on the amplitude, but has a significant effect on the sensitivity, as shown in Figure 8.

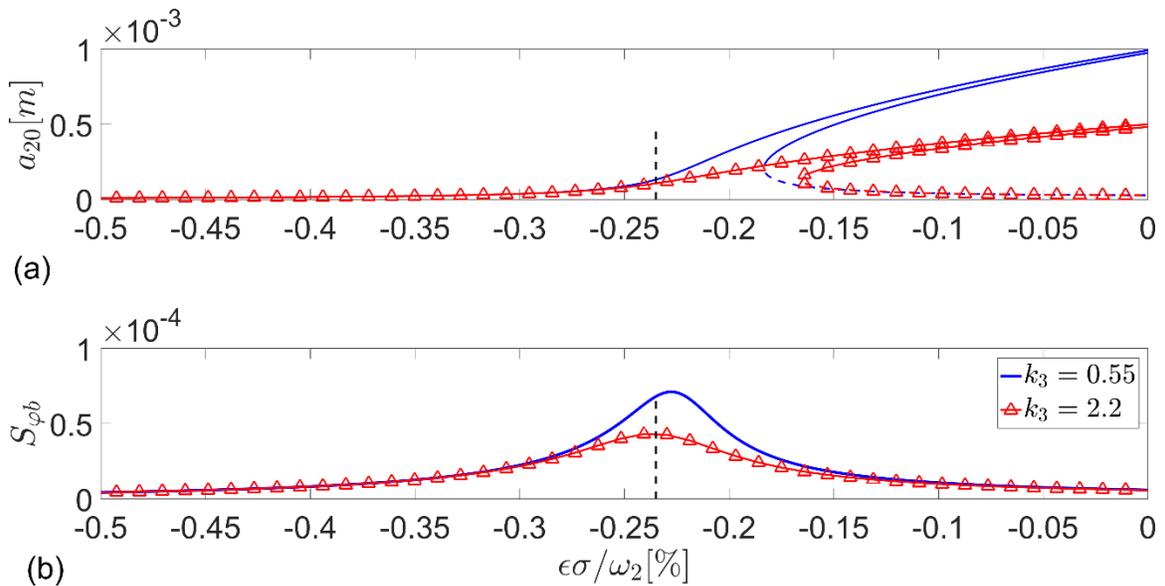

Figure 8: The effect of $k_3$ on the amplitude (top) and sensitivity (bottom). $\sigma_{opt}$ shown by vertical dashed lines. Stable solutions (continuous lines), and unstable solutions (dashed lines).

The response (amplification) and the sensitivity $S_{\varphi b}$ are also very sensitive to parameters which cannot be controlled, such as the amount of imbalance and damping, as shown in Figure 9 and Figure 10.

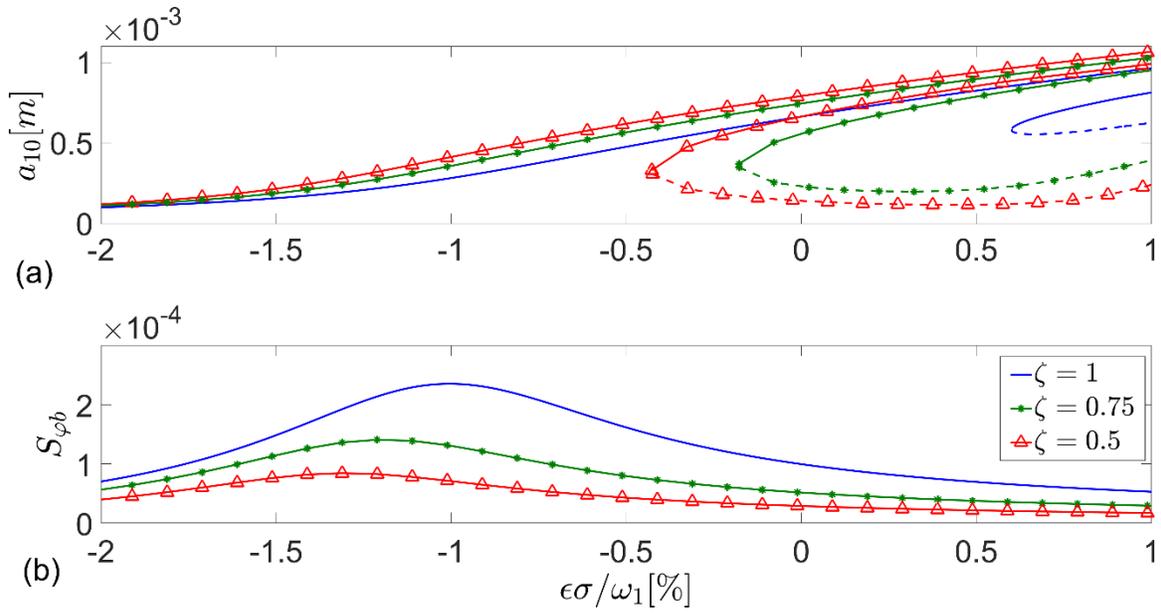

*Figure 9: The effect of $\zeta$ on the amplitude (a) and sensitivity (b), for constant levels of excitation and nonlinearity. Stable solutions (continuous lines), and unstable solutions (dashed lines).*

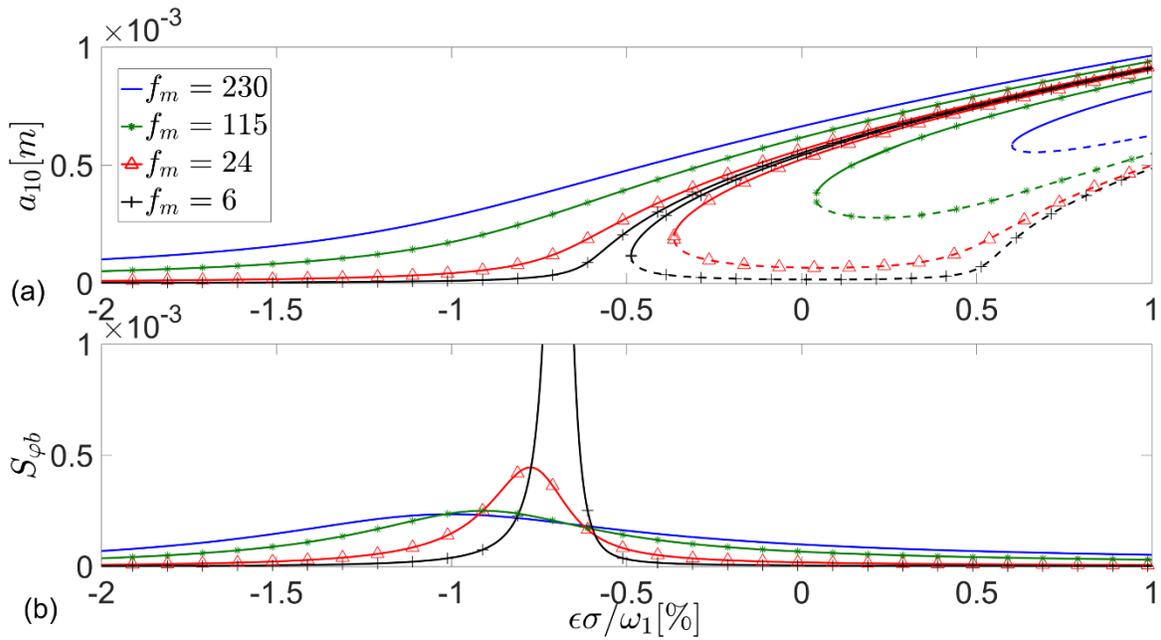

*Figure 10: The effect of modal imbalance $f_m$ (in gr mm) on the amplitude (a) and sensitivity (b). Stable solutions (continuous lines), and unstable solutions (dashed lines).*

The smaller the damping or the imbalance are, the more the frequency response curve becomes narrow banded, bringing the two stable solutions closer to each other, and the sensitivity also becomes more narrow-banded. This means that at low levels of damping and/or imbalance, finding $\sigma_{opt}$ experimentally may become harder, requiring high resolution in the frequency of the excitations.

# 3  Numerical and Experimental Validations

A test rig for the system described in Figure 1 was built, where the controlled forces were applied by linear voice coil actuators. The motion of the plate was measured by laser displacement sensors, and the shaft's speed was controlled by a DC motor, and measured by a magnetic encoder at the shaft's end. The test rig is shown in Figure 11.

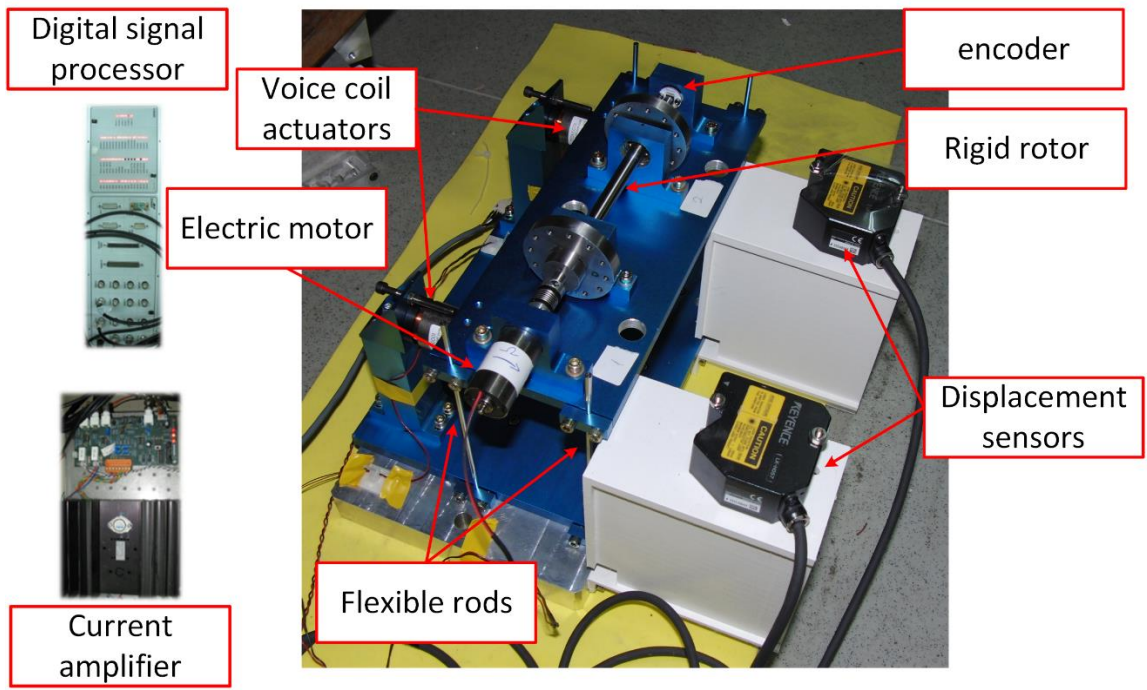

*Figure 11: Test rig*

## 3.1  Calibration process

First, the test rig parameters (natural frequencies, mode shapes and damping ratios) were estimated by means of standard modal testing [36], and the amount of imbalance was estimated by means of standard "Influence Coefficient Method" [5]. The estimated parameters are summarized in Table 1.

*Table 1: estimated system parameters*

| Parameter | Mode 1 | Mode 2 |
|---|---|---|
| Natural frequency $\omega_n$ [Hz] | 18.9 | 29.07 |
| Mode shape $[\phi_{1n} \quad \phi_{2n}]^T$ | $[0.6411 \quad 0.6231]^T$ | $[0.6312 \quad -0.6614]^T$ |
| Damping ratio $\zeta_n$ | 1 | 0.45 |
| Magnitude of modal imbalance $|f_m|$ [gr mm] | 23.9 | 23.9 |
| Phase of modal imbalance $\varphi_j$ [deg] | 260 | 183 |

Second, the proposed method was carried out, without rotating the rotor, but having the voice coils apply known forces that simulate imbalance forces, at a phase angle of 0º. The experiments were conducted at a frequency of 8 Hz (the 1st critical speed was 18.9 [Hz]), with a detuning level of $\sigma/\omega_1 = -1\%$, for which high sensitivity is achieved (see Figure 2, Figure 3 and Figure 15). The results are compared to analytical and numerical calculations, and are shown in Figure 12.

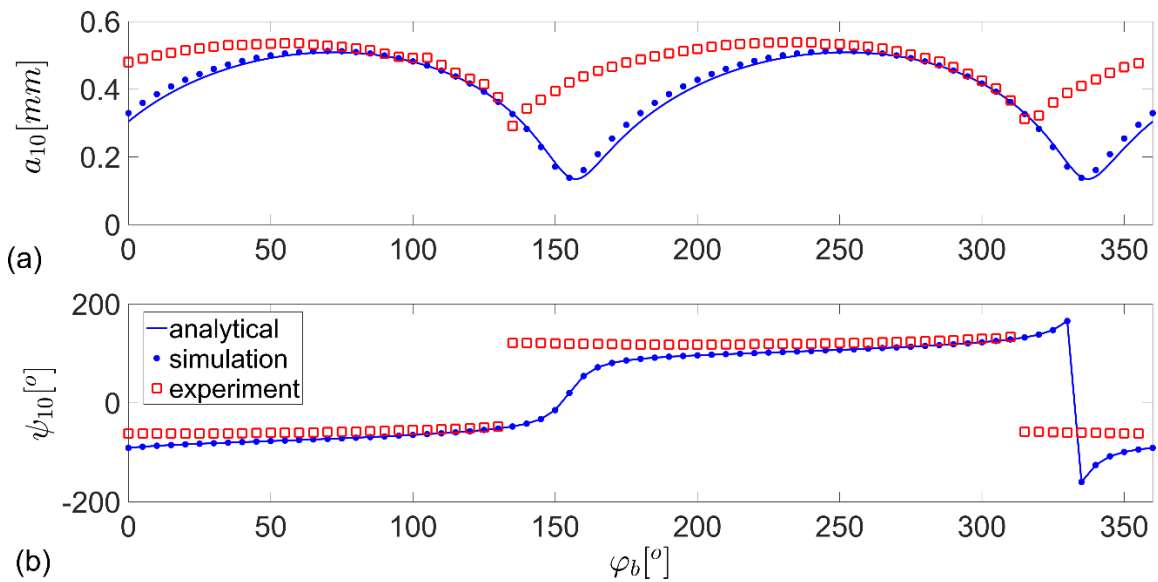

*Figure 12: Calibration process for the 1st mode – sweeping $\varphi_b$. Modal amplitude (a), and phase (b). Analytical solution (lines), numerical simulations (dots) and experimental results (squares).*

As can be seen in Figure 12, the experiments resulted in a periodic response as expected, with a slight shift in the phase with respect to the calculations. The phase of the response, $\psi_{10}$, slightly changes with $\varphi_b$, up to a sharp change of 180º, after reaching $\varphi_{b0}$. At $\varphi_{b0}$, the phase $\psi_{10}$ is 90º away from its value when the amplitude is at its maximum. Since near $\varphi_{b0}$ the phase $\psi_{10}$ changes rapidly it is hard to find its exact value experimentally. Therefore, the value of $\psi_{10}$ at $\varphi_{b0}$ is calculated by the average of the values where the response is maximal, as shown in Table 2. The phase $\varphi_1$ is calculated by Eq. (29), as shown in Table 2.

*Table 2: calibration process results*

|  | **Experiment** | **Analytical** | **Exact** |
|---|---|---|---|
| $\varphi_{b0}$ | 136 / 316 | 157 / 337 | --- |
| $\psi_{10}$ at maximum | -60 / 120 | -72 / 108 | --- |
| $\psi_{10}$ at $\varphi_{b0}$ | 30 / 210 | 18 / 198 | --- |
| $\varphi_1 = -(\psi_{10} + \varphi_{b0})$ | -166 / **14** | 185 / 5 | 0 |

Note that this procedure may lead to a slight error even when the analytical model is used, as shown in Table 2. This is because the evaluation of $\psi_{10}$ at $\varphi_{b0}$ is not exact.

The calibration process shows that the experimental results of the 1st mode need to be corrected by 14º. It is important to note that the numeric simulations are in good agreement with the analytic model (e.g., see Figure 12), which implies that the angle shift is not due to the approximations of the analytic model. The reason for this shift is still unclear, perhaps it is due to the slight error in the estimation of the natural frequency, nonlinearity and/or damping - see Figure 15, or since the dynamics of the electrical system is not modelled.

### 3.2 Balancing the 1st mode

All the experiments were conducted at spin speed of 8 Hz, where the 1st critical speed was 18.9 Hz. The experiments were conducted at the same detuning level used for the calibration process. The initial experiments were conducted with added imbalance, so that $\sigma_{opt}$ would be easier to find (see section 2.3). Since the amount of imbalance is known, the experimental results could be compared to analytical and numerical calculations, as shown throughout this section.

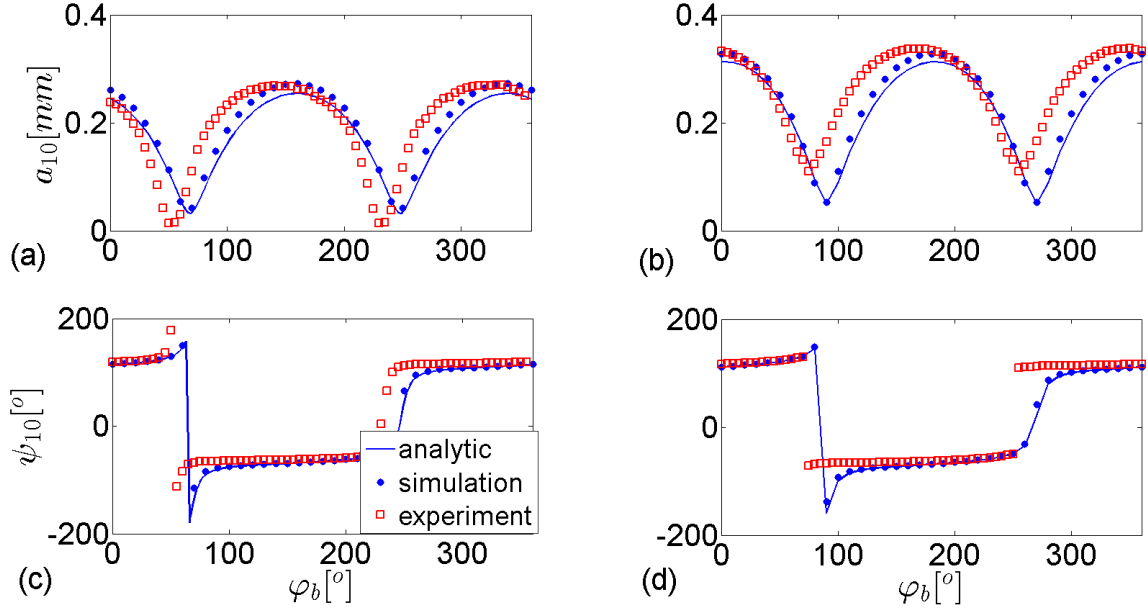

*Figure 13: Sweeping $\varphi_b$. Modal amplitude at run 0 (a); modal amplitude at trial run (b); phase at run 0 (c); phase at trial run (d). Analytical solution (lines), numerical simulations (dots) and experimental results (squares).*

The phase $\varphi_{b0}$ was corrected by 14º, as was found in the aforementioned calibration process. The phase of the imbalance was found according to Eq.(29), as shown in Table 3. A modal trial mass set of 101.7 [gr mm] was placed at 180º (approximately 90º to the location of the modal imbalance, as was calculated from run 0, see Table 3).

*Table 3: extracting phase of modal imbalance for the 1st mode*

|  | Run 0 |  | Trial Run |  |
| --- | --- | --- | --- | --- |
|  | **Experiment** | **Analytic** | **Experiment** | **Analytic** |
| $\varphi_{b0}^{*}$ | 53 / 233 | 67 / 247 | 76 / 256 | 90 / 270 |
| $\varphi_{b0} = \varphi_{b0}^{*} + 14^o$ | 67 / 247 | --- | 90 / 270 | ---- |
| $\psi_{10}$ at maximum | -63 / 117 | -68 / 112 | -64 / 116 | -68.5 / 111.5 |
| $\psi_{10}$ at $\varphi_{b0}$ | 27 | 22 | 26 | 21.5 |

| $\varphi_1$ | **266** / 86 | 271 / 91 | 244 / 64 | 248.5 / 68.5 |

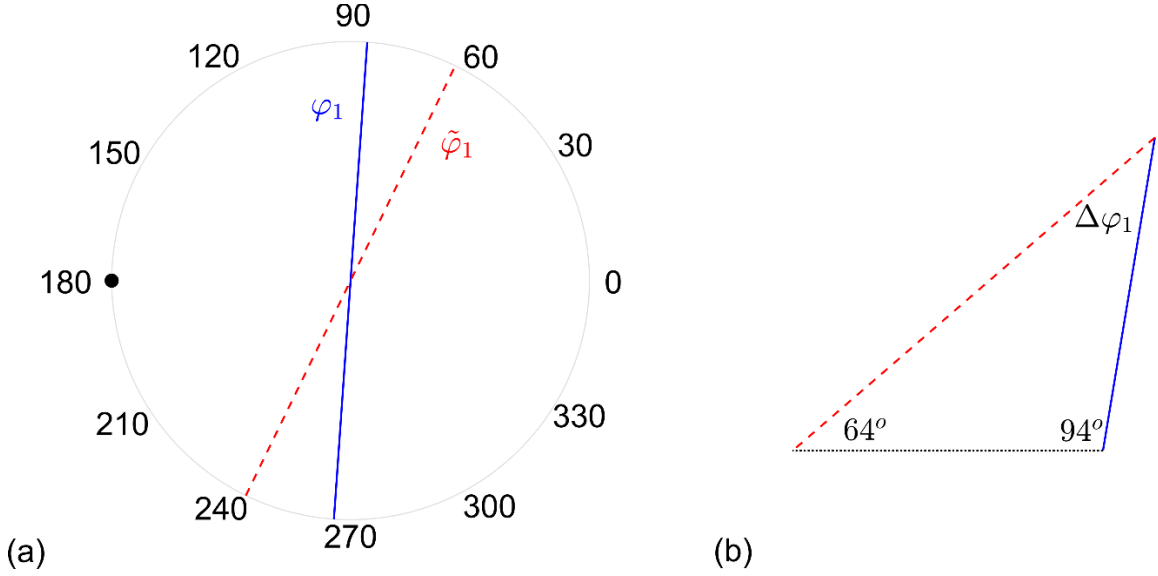

Figure 14: Imbalance calculations for the 1st mode. Possible locations (a); true locations (b). Location at run 0 (blue line), locations at trial run (dashed red line), and modal trial mass (black dot).

Clearly, the imbalance at the trial test must lie between the original imbalance (at run 0) and the trial mass, hence the only possible solution from the experiments is that the original imbalance was at 266°. The magnitude of the original imbalance (at run 0), is found by the rule of sines and is equal to:

$$|f_m| = \frac{\sin(64)}{\sin(22)} 101.7 = 244 \ [gr\ mm] \tag{34}$$

at 266°. The real imbalance (calculated by the influence coefficient method while running close to the critical speed) was: $f_m = 230.7 @ 269°\ [gr\ mm]$. The imbalance was calculated with an accuracy of 92%, where the relative error is calculated by normalizing the residual imbalance by the initial imbalance:

$$error = \left| \frac{244 \cdot \exp\left(i\frac{266 \cdot \pi}{180}\right) - 230.7 \cdot \exp\left(i\frac{269 \cdot \pi}{180}\right)}{230.7} \right| = 7.9\% . \tag{35}$$

### 3.2.1 Smaller amount of imbalance

As was shown in the previous section, the imbalance used was quite large. The frequency response curve which was found experimentally for lower levels of modal imbalance are shown in Figure 15. As can be seen for the case of small imbalance, only past a certain threshold frequency, the measured response agrees with the model, even though the numerical simulations are in excellent agreement with the analytical model at the entire range.

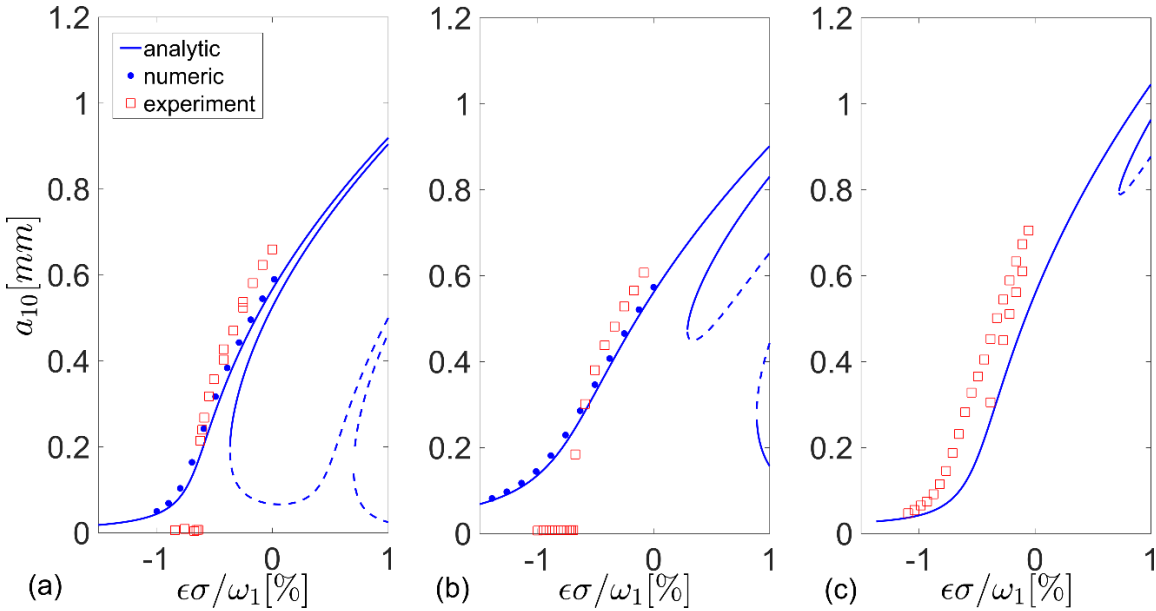

*Figure 15: The effect of the magnitude of the modal imbalance on the frequency response curves. 23.9 [gr mm] (a); 99.3 [gr mm] (b); 230.7 [gr mm] (c). Stable solutions (continuous lines), unstable solutions (dashed lines), numerical simulation results (dots) and experimental results (squares).*

The small differences between the experiments and the analytical model may result from small errors in the evaluations of either the natural frequency, damping ratio or nonlinearity see Figure 8 and Figure 9. It is important to note that in all three cases shown in Figure 15, two stable solutions were found during the experiments, at detuning values for which the analytical model predicts only one stable solution. This probably implies that the damping ratio was lower than the one found during the identification process (see Figure 9).

Sweeping the phase $\varphi_b$ at the lowest detuning level for which amplification was achieved (with a lower amount of imbalance), resulted in switching between the two stable solution branches, as can be seen in Figure 16. The experiments resulted in periodic responses, which depended on the sweep direction, while the intersection between solutions leads to the desired phase $\varphi_{b0}$. This phenomenon was validated by numerical simulations as well as by the analytical model, as shown in Figure 16. Note that the experiments were performed at a detuning level for which the analytical model predicts a single stable solution, as can be seen in Figure 15. In order to achieve two stable solutions in the simulations, the simulations shown in Figure 16 were carried out with a lower level of detuning, for which the analytical model predicts two stable solutions (i.e. $\sigma_{experiment} < \sigma_{simulation} < 0$). Since lower detuning levels lead to higher amplitudes, the magnitude of the response is higher compared to the experiments.

For the smaller amounts of imbalance, as shown in Figure 15, the imbalance was found experimentally with an accuracy greater than 90% in a single iteration.

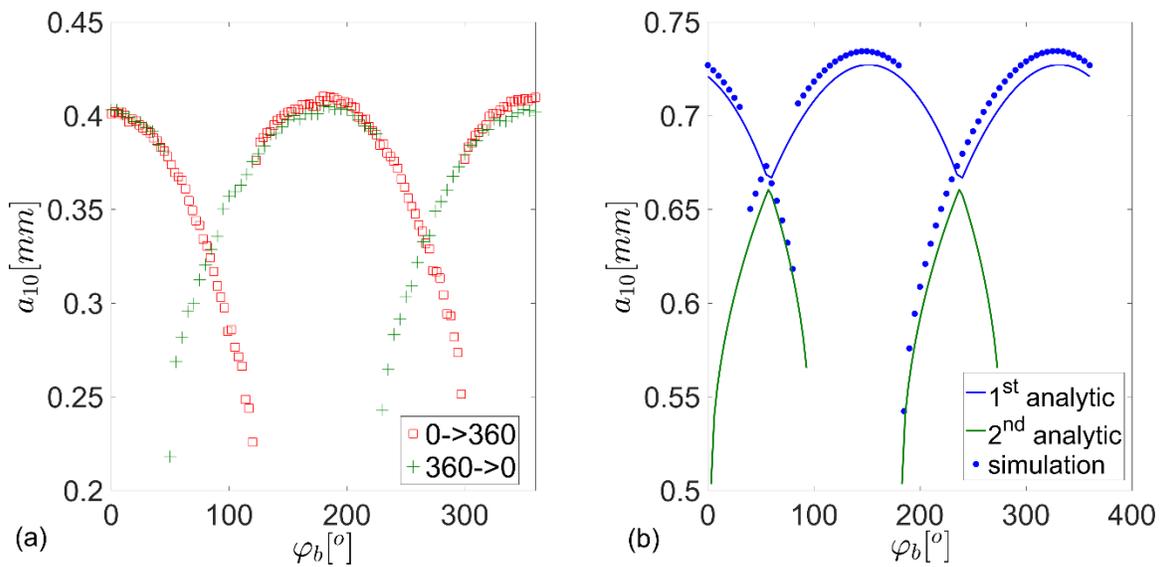

Figure 16: sweeping $\varphi_b$ at multiple solutions zone. Experimental results (a) and calculations (b). Sweeping backwards ('+') and forward (squares). 1st stable analytical solution (blue line), 2nd stable analytical solution (green line) and numerical simulations (dots).

It is important to note that the frequency response curves were found experimentally for $\varphi_b = 0$, however, they depend on $\varphi_b$. For the same value of detuning there are certain values of $\varphi_b$ for which only a single solution exists, and there are values for which two stable solutions exist, as shown in Figure 16.

The reason for lack of amplification at larger values of detuning, where the sensitivity to the phase $\varphi_b$ is higher is not clear. Although increasing the values of parametric excitations $(k_{pa}, k_{pb})$ slightly, seems to be the solution, it did not yield amplification for the entire range of $\varphi_b$. It should be mentioned that the numerical simulations are in excellent agreement with the analytical model. In previous work [37], numerical simulations of the test rig for very low imbalance (24.5 and 14.2 [gr.mm] for the 1st and 2nd mode, respectively) resulted in significant amplification for every value of $\varphi_b$. Furthermore, the simulations in [37] were carried out for a spin velocity of 3.5 Hz, so the modal forces of the 1st mode were less than 20% comparing to the lowest value of imbalance used in the experiments shown in Figure 15.

### 3.3 Balancing the 2nd mode

All experiments were conducted at spin speed of 8 Hz, where the 2nd critical speed is 29 Hz. The tests were conducted at detuning of $\sigma/\omega_2 = -0.44\%$, which was the maximum detuning level for which amplification was achieved, see Figure 17. The same detuning level was used for the calibration process, which resulted in a correction of 7°, see Figure 18 and Table 4. The experimental results of sweeping the phase $\varphi_b$ are shown in Figure 19:, and the calculation of the imbalance is shown in Table 5 and Figure 20.

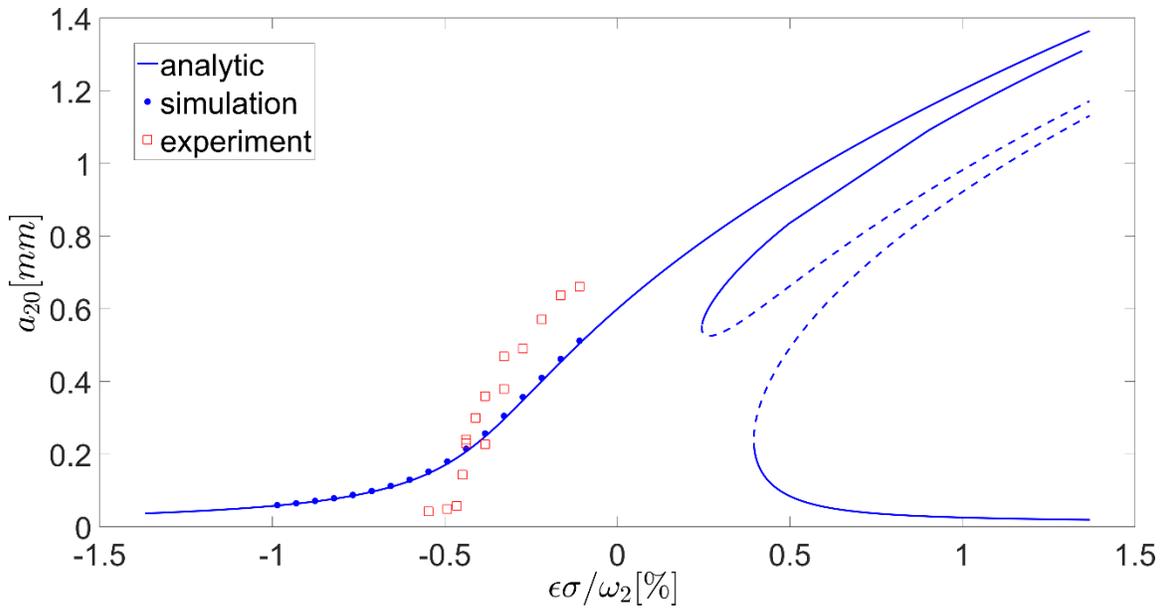

*Figure 17: Frequency response curve for the 2nd mode. Analytical (lines), numerical (dots) and experimental results (squares). Stable solutions (continuous lines), and unstable solutions (dashed lines).*

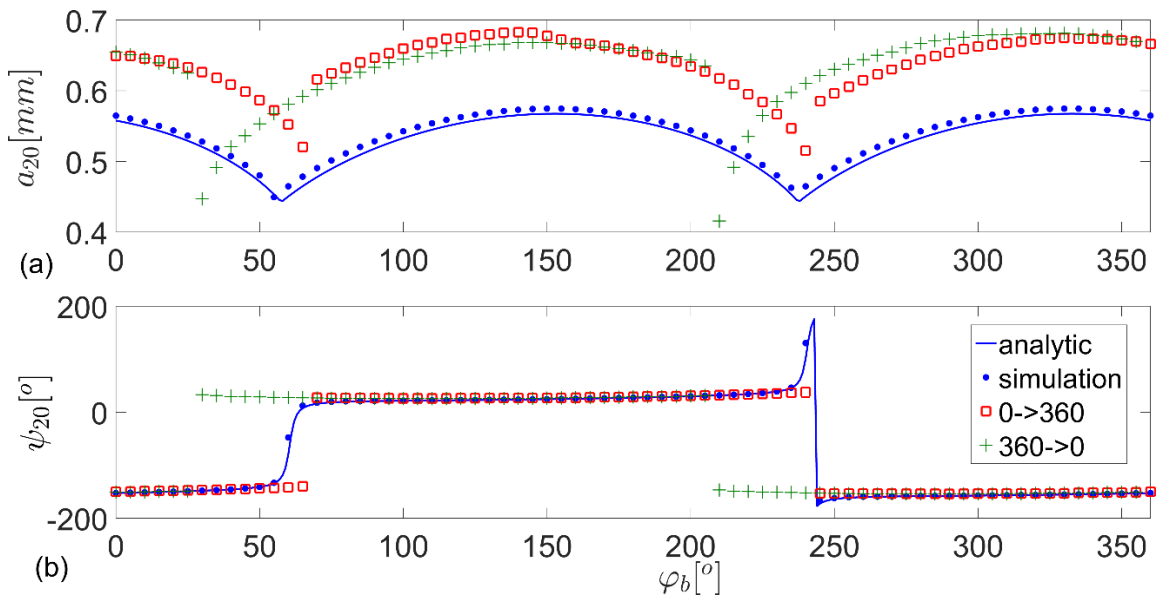

*Figure 18: Calibration process for the 2nd mode – sweeping $\varphi_b$. Modal amplitude (a), and phase (b). Analytical solution (lines), numeric simulations (dots) and experimental results (squares and '+').*

Table 4: Calculation of calibration value for the 2nd mode

|  | Experiment | Analytic | Exact |
|---|---|---|---|
| $\varphi_{b0}$ | 55.5 / 228 | 60.5 / 240.5 | --- |
| $\psi_{20}$ at maximum | 27 / 206.5 | 25 / 205 | --- |
| $\psi_{20}$ at $\varphi_{b0}$ | 117 | 115 | --- |
| $\varphi_2 = -(\psi_{20} + \varphi_{b0})$ | 187.5 / 15 | 184.5 / 4.5 | 0 |

Note that in this case the two values of $\varphi_{b0}$ (and hence also $\varphi_2$) which were found experimentally are not exactly 180° apart. One of the values is 3° away from the analytical solution, while the other is 10.5°. We took the average of those deviations, so the required calibration for the 2nd mode is 7°. A modal trial mass set of 204.8 [gr.mm] was placed at 270° (approximately 90° to the location of the modal imbalance, as was calculated from run 0, see Table 5).

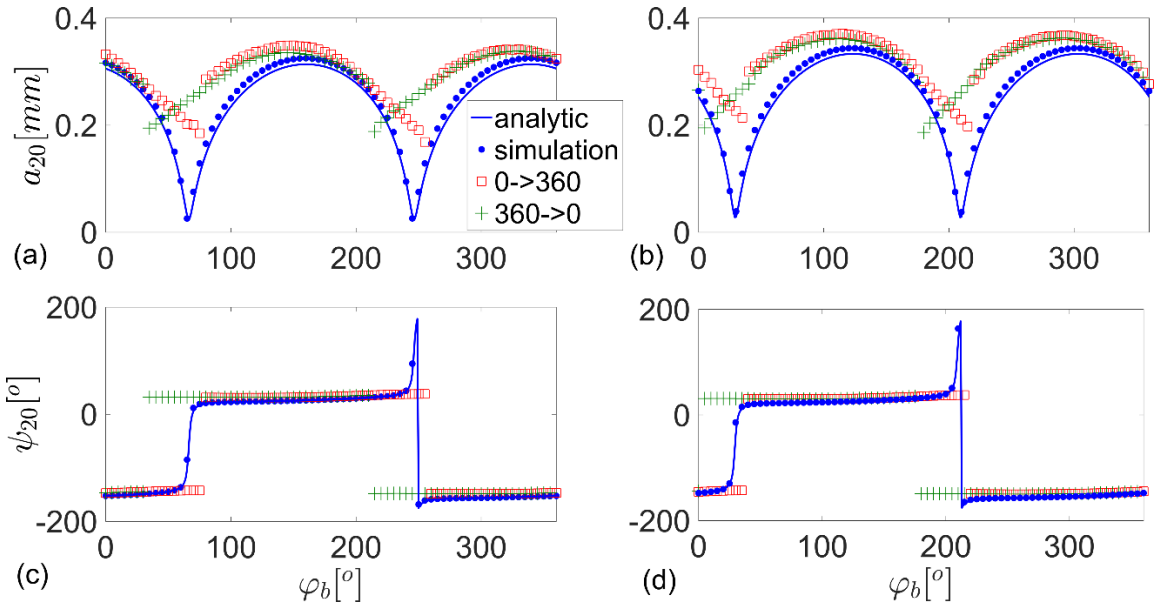

Figure 19: Sweeping $\varphi_b$. Modal amplitude at run 0 (a); modal amplitude at trial run (b); phase at run 0 (c); phase at trial run (d). Analytical solution (lines), numerical simulations (dots) and experimental results (squares and '+').

Table 5: calculation of the modal imbalance's phase for the 2nd mode

|  | Run 0 | | Trial Run | |
| --- | --- | --- | --- | --- |
|  | Experiment | Analytic | Experiment | Analytic |
| $\varphi_{b0}^{*}$ | 54 / 232 | 67.5 / 247.5 | 22 / 199 | 29.5 / 209.5 |
| $\varphi_{b0}^{*} = \varphi_{b0} + 7^{o}$ | 61 / 239 | --- | 29 / 206 | --- |
| $\psi_{20}$ at maximum | 32 / 212 | 26 / 206 | 30.5 / 210.5 | 25 / 205 |
| $\psi_{20}$ at $\varphi_{b0}$ | 122 | 116 | 120.5 | 115 |
| $\varphi_{2}^{*}$ | 177 / 359 | 176.5/356.5 | 210.5/ 33.5 | 215.5 / 35.5 |
| $\varphi_{2}$ | 178 /358 | --- | 212 /32 | --- |

Once again, the experimental values of $\varphi_{b0}$ are not exactly 180º apart. We referred to the average so the possible values of $\varphi_{2}$ are 178º /358º for run 0, and 212º /32º for the trial run.

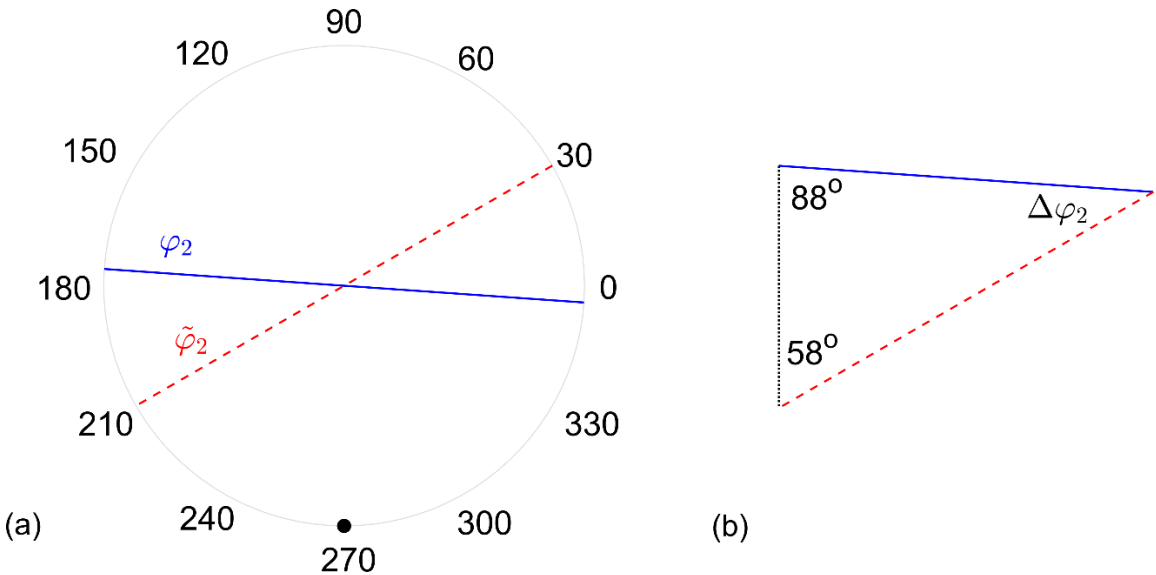

Figure 20: Imbalance calculation for the 2nd mode. Possible locations (a); true locations (b). Location at run 0 (blue line), locations at trial run (dashed red line), and modal trial mass (black dot).

Similarly to the 1st mode, the imbalance at the trial test must lie between the original imbalance (at run 0) and the trial mass, hence the only possible solution from the experiments is that the original imbalance was located at 176°, and is of size:

$$|f_m| = \frac{\sin(58)}{\sin(34)} 204.8 = 310.6 \ [gr\ mm]$$

The real imbalance (calculated by the influence coefficient method, and the imbalance weigths) was: $f_m = 291.6\ @\ 174^o\ [gr\ mm]$. The imbalance was calculated with an accuracy of 90%.

## 4 Conclusions

The current paper presents a novel balancing procedure, outlining a balancing procedure of high frequency modes, while running at low speeds, and where the response to imbalance is very low. The procedure is based on the non-degenerate parametric amplifier shown in [30,31]. The parametric amplifier significantly amplifies the response of the system at any desired mode, while keeping the response dependant on the projection of the imbalance on the desired mode. By applying the parametric excitation orthogonally to undesired modes, the parametric amplifier enables the detection of the projection of the imbalance on any desired mode (phase and magnitude).

The method is based on two runs of the rotor, an initial run, and a run with a trial mass, from which the imbalance is found. In each run, the phase of the excitation is changed until reaching minimum amplitude of vibration. It is also possible to use the actuators to simulate imbalance forces instead of applying trial masses.

Numerical simulations and initial experiments validated the analytical model. The method was proved to provide substantial amplification of the response to imbalance, while rotating well below the critical speeds. The estimation of the imbalance was shown to be accurate by at least 90% in a single balancing iteration, for various values of imbalance, for both modes.

Initial experiments showed that a calibration procedure is required to compensate for an offset between the experimental results and theory, perhaps due to small errors in the estimated parameters such as the natural frequencies, damping ratios and the nonlinearity.

In the initial experiments for low levels of imbalance, there was no amplification at high detuning levels. Although this did not prevent the method from working, it required working in detuning levels for which two stable solutions exist, which required the phase to be swept back and forth. The reason for the lack of amplification is not yet understood, and will be examined in future work.

## Appendix A: Non-diagonal $\mathbf{K_p}(t)$

If the parametric excitations are not orthogonal to the undesired mode (i.e., if the matrix $\mathbf{K_p}(t)$ is not diagonal), both modes tend to be excited due to the "blending element". This section deals with the case of a single actuator, which may be a reason for not having an ideal excitation where $\mathbf{K_p}(t)$ is diagonal. The modal amplitude of the $k^{th}$ mode (where the $k^{th}$ mode is excited), for excitation at the $j^{th}$ plane is:

$$a_k = \frac{k_{pb}\phi_{jk}\left[\phi_{j1}\Lambda_1 \sin(\psi + \varphi_1 + \varphi_b) + \phi_{j2}\Lambda_2 \sin(\psi + \varphi_2 + \varphi_b)\right]}{\phi_{jk}^2 k_{pa} \sin(2\psi + \varphi_a) - 4\omega_k^2 \zeta_k} \quad (A1)$$

In this case, the response depends on the projection of the imbalance on both modes, due to the "blending element". The response in this case is not linear with respect to the projection of imbalance on the desired mode. Furthermore, the response will not reach a minimum according to Eq. (29). However, there is a phase $\varphi_{b0}$ which leads to a minimum response (except for the case where $\phi_1 = \phi_2, \Lambda_1 \neq \Lambda_2$).

In this case, there are four unknowns: $\Lambda_1$, $\Lambda_2$, $\varphi_1$ and $\varphi_2$ (or alternatively, two imbalance masses and phases at the balancing planes). Finding the four unknowns requires at least four runs with different (known) trial masses at each run. Each run includes changing the value of

$\varphi_b$, until $\varphi_{b0}$ is reached, hence a nonlinear equation system is constructed, which can be solved numerically to find the four unknowns.

## Appendix B: Stability Calculation

The stability of the solution, is found by the following eigenvalue problem [35]:

$$\begin{bmatrix} \dfrac{\partial a'_j}{\partial a_{j1}} & \dfrac{\partial a'_j}{\partial \psi_{j1}} \\ \dfrac{\partial \psi'_j}{\partial a_{j1}} & \dfrac{\partial \psi'_j}{\partial \psi_{j1}} \end{bmatrix} \boldsymbol{\theta} = \lambda \boldsymbol{\theta} \tag{B1}$$

where $a_j = a_{j0} + a_{j1}$, $\psi_j = \psi_{j0} + \psi_{j1}$, and $a_{j1}, \psi_{j1} \ll 1$, and $a_j{}', \psi_j{}'$ are shown in Eq. (27). The steady state motions are stable if both roots, $\lambda$, have negative real parts.